\documentstyle[11pt,newpasp,twoside,epsf]{article}
\def\edcomment#1{\iffalse\marginpar{\raggedright\sl#1\/}\else\relax\fi}
\reversemarginpar

\begin{document}

\title{Asteroseismology for the Masses}

\author{Travis S. Metcalfe}

\affil{Department of Astronomy, Mail code C1400, 
       University of Texas, Austin, TX 78712    }

\begin{abstract} 
For many years, astronomers interested in white dwarf stars have promised
that the study of these relatively simple compact objects would ultimately
lead to useful information about the physics under extreme conditions of
temperature and pressure. We are finally ready to make good on that
promise. Using the observational techniques of the Whole Earth Telescope
developed over the past decade, and a new analytical method only recently
made feasible by the availability of fast inexpensive computers, we
demonstrate that meaningful constraints can now be made on the rates of
important nuclear fusion reactions which cannot presently be measured in
the laboratory.
\end{abstract}

\section{Introduction}

The Whole Earth Telescope (WET) is an informal collaboration between
astronomers around the world who monitor pulsating white dwarfs
continuously for up to two weeks to avoid the aliasing problems inherent 
in single-site data and to resolve the many frequencies present in these
stars (Nather et al.~1990). The WET first observed the helium-atmosphere 
variable (DBV) white dwarf GD~358 in 1990 (Winget et al.~1994) and found 
a series of nearly equally-spaced periods in the Fourier Transform which 
they interpreted as non-radial g-mode pulsations of consecutive radial 
overtone.

Bradley \& Winget (1994) attempted to match theoretical models to the 11
identified periods and to the period spacing. Their best-fit model had a thin 
helium layer, about $10^{-6}$ times the mass of the star, and this posed 
a problem for theory which still hasn't been adequately resolved.
Computational resources have improved significantly since the original 
fit, so we decided to explore a larger region of parameter-space than was 
possible at the time. Our goal was to establish a method for fitting 
pulsation models to the data that is both more global, and more objective.

\section{Computational Method}

\subsection{Model Parameters}

Aside from the core composition, the three most important parameters that
determine the pulsation characteristics of DBV white dwarf models are the
stellar mass, the temperature, and the helium layer mass. 

The observed mass distribution of white dwarfs is strongly peaked near
0.6~$M_{\sun}$ with a FWHM of about 0.1~$M_{\sun}$ (Napiwotzki Green \& 
Saffer 1999). We decided to explore a range of masses from 0.45 (the 
theoretical cutoff for C/O white dwarfs) to 0.95~$M_{\sun}$. 

The most recent temperature determinations for the eight known DBV stars,
depending on various assumptions, place the lowest temperature pulsator
at about 22,000~K and the highest temperature pulsator at nearly 28,000~K
(Beauchamp et al.~1999). We decided to explore all temperatures between 
20,000 and 30,000~K.

As for the helium layer, above a fractional mass of about $10^{-2}$,
helium will start to burn at the base of the envelope. Thinner than about
$10^{-8}$ and our models no longer pulsate (Bradley 1993). We didn't go 
quite this thin in order to keep our models running in a fully automated 
way, for reasons which will become clear later.

To give you a sense of how much more parameter-space we are covering,
Figure 1 shows a front and side view of the 3-dimensional search space
we've defined. The dashed boxes show the range of parameters considered by
Bradley \& Winget, while the solid boxes show the range we are considering.
Our search space has more than $100\times$ the volume and a much finer 
resolution (100 points in each dimension).

\begin{figure}
\plotone{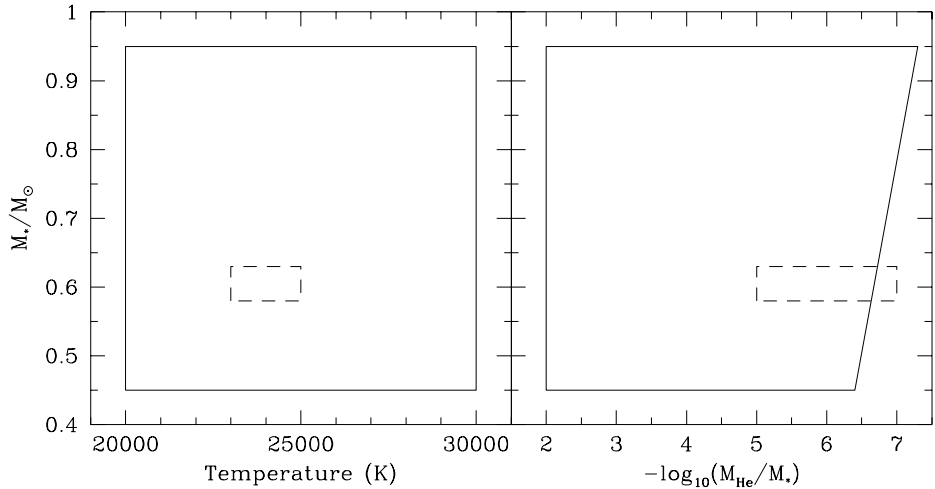}
\caption{The relative sizes of the white dwarf model parameter-space 
searched by Bradley \& Winget (dashed line) and by this study (solid line).}
\end{figure}

At this point, we clearly have an approach that is more global. Now we
could just calculate all of these models and compare them to the data, and
we would get an objective solution. But we don't have to do that---there's 
a better, more efficient way.

\subsection{Genetic Algorithms}

A genetic algorithm (GA) uses an optimization method based on an analogy with
biological evolution. You can think of GAs as a sort of iterative Monte
Carlo method directed by a memory of what worked well in the past, and what 
didn't work so well. For our white dwarf models, this method can actually 
find the global minimum with about 1/10th the number of model evaluations 
that would be required to calculate the full grid. This is better, but it's 
still pretty computationally intensive, so we built a specialized 
computational instrument just for the job (see Figure 2).

\begin{figure}
\plotone{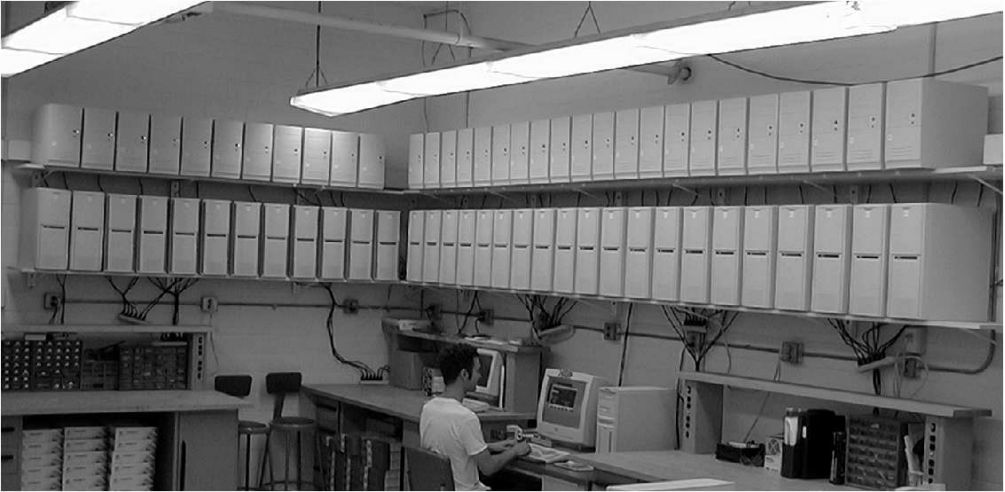}
\caption{The asteroseismology metacomputer, a collection of 64 minimal
PCs connected by a network, which we used to perform the calculations
in parallel.}
\end{figure}

We call this new instrument {\it Darwin}. It consists of 64 minimal PCs 
connected by a network, and can compute our models in parallel with a 
speedup factor of about 60 (see Metcalfe \& Nather 1999a,b). Each box
contains only a processor, 32 Mb of memory, and an ethernet card with a 
custom Linux bootrom.

\subsection{Proof of Principle}

Once our code was running on this machine, the first thing we did was 
generate a set of pulsation periods from a model within the search space, 
and then let the GA try to find the input parameters.

In 9 out of 20 runs using different random number sequences, the GA found
the exact answer. In an additional 4 runs, it ended up with an answer
close enough to the correct one for a small grid to yield the exact
answer. This means that for a given run, there is a 35\% chance that we
won't find the right answer, but if we perform 5 independent runs there 
will be a 99\% probability of finding the right answer. 

Next, we added noise to the input model to double check that the 5 run 
method was sufficient. We estimated the level of noise present on the 
observed frequencies in two ways. First, we used the identifications of 
the 63 linear combination frequencies (Vuille et al.~2000) to determine 
the distribution of differences between the observed and predicted 
frequencies. The best-fit Gaussian to this distribution had 
$\sigma=0.17~\mu$Hz. Second, we passed 100 synthetic light curves 
through a simulation of the standard WET reduction procedure, and looked 
at the distribution of differences between the input and output 
frequencies. This distribution was slightly narrower, having a best-fit 
Gaussian with $\sigma=0.053~\mu$Hz. We ran the GA method on several 
realizations of each estimate of the noise, and in all cases it found 
the right answer.

\section{Results}

Finally, we applied the new method to the real data for GD~358. To 
facilitate comparison with earlier work we ran the GA on six different 
combinations of core composition and transition profiles: pure carbon, 
pure oxygen, and mixed 50:50 \& 20:80 C/O cores with both ``steep'' and 
``shallow'' transition profiles (see Bradley Winget \& Wood 1993). 

The first thing we noticed was that there were two families of solutions
in the larger search space, corresponding to thick and thin helium layers. 
These two families did not contain equally good solutions---one was always
clearly better than the other---but which family was preferred depended on
the core composition. Notably, when we restricted the GA to looking in the 
region of parameter-space searched by Bradley \& Winget, it found the same 
answer that they found.

In general, reasonably good fits were possible for every core composition, 
but excellent fits were only possible in one specific case: the 20:80 C/O 
``shallow'' mix. This new best-fit model is shown in the top panel of 
Figure 3 with the solution of Bradley \& Winget shown in the bottom panel 
for comparison. The root-mean-square (r.m.s.) period difference between 
the new best-fit model and the observations is 1.50 seconds, compared to 
2.69 seconds for Bradley \& Winget's solution. The second best fit found 
by the GA had an r.m.s.~of 1.76 seconds---significantly worse considering 
that the noise on the individual periods amounts to only a few hundredths 
of a second.

\begin{figure}
\plotone{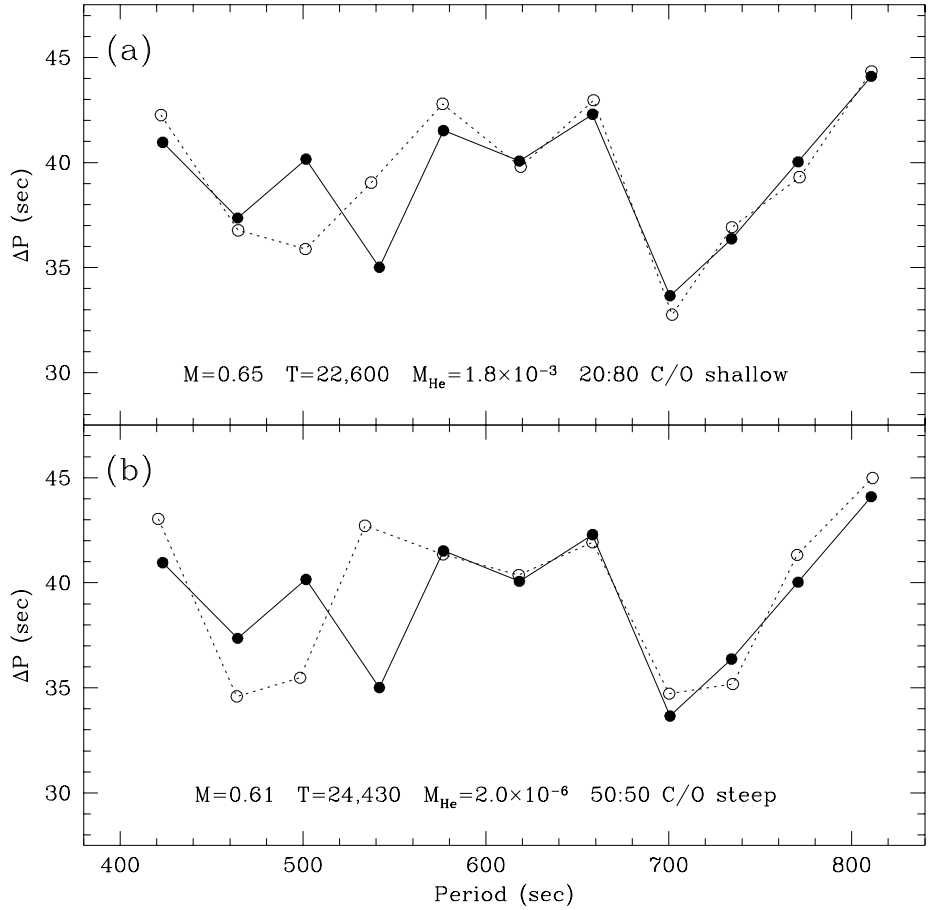}
\caption{The observed periods and period-spacing for GD~358 (solid points) 
with the best-fit models (open points) from (a) the genetic algorithm,
and (b) Bradley \& Winget (1994).}
\end{figure}

\section{Future Work}

Although the new best-fit model is significantly better than the previous 
solution, it clearly still needs improvement. It's possible that the 
duality between thick and thin helium layers for GD~358 is an indication
that composition transitions exist in {\it both} places. This idea has
been followed up in the context of possible $^3$He diffusion by
Montgomery Metcalfe \& Winget in these proceedings.

Another possibility is that more realistic C/O transition profiles, such as
those of Salaris et al.~(1997), will provide the extra structure needed by
the models to fit the observed data. Combining the use of these profiles
with a treatment of the central oxygen abundance as a free parameter by the
GA, we can realistically expect to measure the central C/O ratio in GD~358. 
During helium burning the $^{12}{\rm C}(\alpha,\gamma)^{16}{\rm O}$ and the 
$3\alpha$ reactions are competing for the available $\alpha$-particles, so
the final carbon to oxygen ratio in the core provides a direct measure of 
the relative rates of these two reactions (Buchman 1996). This is 
especially significant considering the difficulty of measuring the 
$^{12}{\rm C}(\alpha,\gamma)^{16}{\rm O}$ reaction in the laboratory.

Finally, we may be able to reverse-engineer the shape of the C/O transition
profile by parameterizing the Brunt-V\"ais\"al\"a frequency and investigating
what changes to the shape yield significantly better fits to the data. This
would provide an independent check on the theoretical profiles, and would
open up an entirely new approach to improving our models.

\acknowledgements I would like to thank Ed Nather and Don Winget for their
guidance; Mike Montgomery, Paul Bradley, S.O. Kepler and Craig Wheeler for
helpful discussions; Paul Charbonneau for supplying an unreleased version
of the PIKAIA genetic algorithm; and Gary Hansen for arranging the donation
of 32 computer processors through AMD.

\end{document}